\documentclass[useAMS,usenatbib]{mn2e}

\usepackage{graphicx}
\usepackage{dcolumn}
\usepackage{bm}
\usepackage{amsmath}
\usepackage{amsfonts}
\usepackage{amssymb}
\usepackage{psfrag}
\usepackage{here}
\usepackage{rotating}
\usepackage{fixltx2e}
\usepackage{float}

\title[Blazar Lorentz factors from optical Fundamental Plane]{Lorentz factor distribution of blazars from the optical Fundamental plane of black hole activity}
\author[P. Saikia et al. 2016]{Payaswini Saikia$^{1}$\thanks{E-mail:
p.saikia@astro.ru.nl}, Elmar K\"{o}rding$^{1}$ and Heino Falcke$^{1,2}$\\
$^{1}$Department of Astrophysics/IMAPP, Radboud University, Nijmegen, P.O. Box 9010, 6500 GL Nijmegen, The Netherlands\\
$^{2}$ASTRON, Oude Hoogeveensedijk 4, 7991 PD Dwingeloo, The Netherlands}
\begin{document}

\pagerange{\pageref{firstpage}--\pageref{lastpage}} \pubyear{2016}

\maketitle

\label{firstpage}


\begin{abstract}

\noindent Blazar radiation is dominated by a relativistic jet which can be modeled at first approximation using just two intrinsic parameters - the Lorentz factor $\Gamma$ and the viewing angle $\theta$. Blazar jet observations are often beamed due to relativistic effects, complicating the understanding of these intrinsic properties. The most common way to estimate blazar Lorentz factors needs the estimation of apparent jet speeds and Doppler beaming factors. We present a new and independent method of constructing the blazar Lorentz factor distribution, using the optical fundamental plane of black hole activity.  The optical fundamental plane is a plane stretched out by both the supermassive black holes and the X-ray binaries, in the 3D space provided by their [OIII] line luminosity, radio luminosity and black hole mass. We use the intrinsic radio luminosity obtained from the optical fundamental plane to constrain the  boosting parameters of the VLBA Imaging and Polarimetry Survey (VIPS) blazar sample. We find a blazar bulk Lorentz factor distribution in the form of a power law as $N(\Gamma) \propto \Gamma^{-2.1 \pm 0.4}$ for the $\Gamma$ range of 1 to 40. We also discuss the viewing angle distribution of the blazars and the dependence of our results on the input parameters.

\end{abstract}



\section{Introduction}

\indent Blazars, including both BL Lac objects and flat-spectrum radio quasars (FSRQs), represent the most powerful and highly variable class of active galactic nuclei (AGNs), with flat radio spectra, core-dominated radio morphology and radiation dominated by a relativistic jet oriented close to our line of sight \citep[eg.][etc.]{bk79, up95}. Blazar jets can be strongly beamed depending on the relativistic Doppler factor of the source. The relativistic jet of a blazar, at first approximation, can be defined by just two intrinsic parameters - the Lorentz factor $\Gamma$ and the viewing angle $\theta$. Studying the radio jets of these sources is crucial to understand the kinematics of black holes and the laws of physics under extreme conditions. Specially, having information on the fundamental and intrinsic properties of the relativistic jets, like Lorentz factors and viewing angles, will help to constrain the physics of jet launching region, shed light in the formation, collimation, acceleration, and propagation of the jets, kinematically characterize the sources and estimate the underlying intrinsic physical properties of the sources, like their luminosity function \citep[eg.][]{a12}. But owing to difficulties involved in measuring the intrinsic luminosity of the blazar jet, not much is known regarding these intrinsic parameters of blazars, neither individually nor as a population. In this paper, we present a novel method to obtain the distribution of these parameters for a population of blazars.

The most common way to calculate the blazar Lorentz factor is by observing the apparent speed of the jet and estimating the Doppler beaming factor. The apparent speed of the jets can be found directly by using Very Long Baseline Interferometry (VLBI) observations \citep[eg.][etc.]{j01,k04,b08}. Estimating the Doppler beaming factor is more complicated and \cite{lv99} has compared the different methods currently used to estimate it and have found that a typical radio-loud quasar has a Lorentz factor $\sim10$ and a viewing angle $<5^{\circ}$, while a typical BL Lac object has a Lorentz factor $\sim 5$ and a viewing angle $< 10^{\circ}$.

Doppler beaming factors can be estimated by calculating the decline of flux with time of a jet component and comparing it to the measured size of the VLBI component \citep[eg.][etc.]{j05}; by combining X-ray observations with radio fluxes at turnover frequencies \citep[eg.][etc.]{g93,b07} or by observing the brightness temperature of the source and comparing it to the equipartition temperature \citep[eg.][etc.]{r94,h09}.

But all these methods have their own assumptions and limitations. Using the decline of flux of a jet with time to find Doppler factor needs the assumption that the variability timescale of a resolved jet component is determined by the light travel time across the component, rather than loss processes and relies upon the observational determination of angular sizes of the component and the variability timescale \citep{j05}. Accurate measurements of Doppler factors by combining X-ray observations with radio fluxes require simultaneous X-ray and radio data taken at the turnover frequency, without which large errors are induced; while on the other hand using the brightness temperature method to obtain Doppler factors is advantageous as it needs single-epoch radio observations, but the values again need to be obtained at the turnover frequency \citep{lv99}.

These limitations are difficult to overcome, if the Doppler boosting factor of a blazar is being estimated on a source-by-source basis, but it is possible to explore the connection between observed and intrinsic properties for a complete sample of blazars and find a Lorentz factor distribution for the blazars as a population. The distribution for Lorentz factors has been explored by many studies \citep[eg.][etc.]{pu92,lm97,k04,j05,a12}. \cite{f95} used the spread around the quasar radio/optical correlation to constrain the jet Lorentz factors between 3 and 10. \cite{pu92} compared the luminosity functions of flat spectrum radio quasars with relativistic models and found very good agreement for a distribution of Lorentz factors $5 \lesssim \Gamma \lesssim 40$, distributed as $N(\Gamma) \propto \Gamma^{-2.3}$. \cite{lm97} has used the Caltech-Jodrell Bank sample of bright, flat-spectrum, radio core-dominated AGNs (CJ-F) and found that their model predictions are consistent with the CJ-F data for a parent Lorentz factor distribution of the form $N(\Gamma) \propto \Gamma^{\alpha}$, where $-1.5 \lesssim \alpha \lesssim -1.75$. 

In this paper we present a new, independent method to estimate blazar Lorentz factor and viewing angle distributions by using the optical fundamental plane of black hole activity. The fundamental plane (FP) is a relation connecting hard state X-ray binaries (XRBs) and their supermassive analogs, suggesting that black holes regulate their output similarly across the entire mass scale \citep{m03, f04}. Observationally, it is a plane stretched out by black holes over the entire mass range, in the 3D space provided by their X-ray luminosity, radio luminosity and black hole mass. It is given by the relation $ \log L_{R} = 0.60 \log L_{X}  + 0.78 \log M$, where $L_{R}$ is the core radio luminosity at 5 GHz, $L_{X}$ is the 2-10 keV X-ray luminosity and $M$ is the central black hole mass. These scalings can be well explained by the theory of synchrotron emitting compact relativistic jets and radiatively inefficient accretion flows \citep{bk79,fb95,hs03}. Depending on the AGN sample, the regression methods used to derive the FP coefficients can result in a significant scatter around the correlation, from $\sim$ 0.38 - 0.65 dex \citep{k06}, specially if one includes high-state objects, which most likely have additional emission components and hence can introduce a higher uncertainty.

The FP has been re-established in the optical band, using the forbidden [OIII] emission line luminosity as an indirect tracer of accretion rate instead of X-ray luminosity \citep{s15}. The optical FP is a very useful tool to study the intrinsic properties of blazars as the [OIII] emission line luminosity is direction-independent and is not affected by relativistic boosting, while both radio and x-ray luminosities can have significant contribution from the relativistic jet and hence have beaming effects. Moreover, blazar sources, specially the low-frequency peaked BL Lacs can have the synchrotron cut-off of the jet spectrum below the X-ray band. For such sources, the observed X-ray luminosity would be dominated by synchrotron self-Compton emission and hence the [OIII] line luminosity is a much better proxy of the accretion rate.

With the optical FP relation, we can use the black hole mass and the [OIII] line luminosity to estimate radio luminosity, i.e. the intrinsic power of the jet. This plane relation is given as $\log L_{R} = 0.83 \log L_{OIII}  + 0.82 \log M$ using supermassive black holes (SMBH) only, while including the X-ray binaries in the analysis to reproduce the fundamental plane in X-rays has resulted in the relation $\log L_{R} = 0.64 \log L_{X}  + 0.61 \log M$. The radio and the OIII fluxes used in formulating this relation are directly observed, while the mass of the black holes are inferred from stellar velocity dispersions. We discuss the implication of using an inferred quantity to calculate the intrinsic radio luminosity in Section 2.2. For the plane obtained with the supermassive black holes, an intrinsic scatter of 0.35 dex was found, which was reduced to 0.2 dex after including the X-ray binaries in the plane.

We use intrinsic radio luminosities obtained from the optical fundamental plane relation to estimate the boosting parameters of flat-spectrum radio sources from the VLBA Imaging and Polarimetry Survey (VIPS), consisting of 1,127 sources with flat radio spectra taken from the Cosmic Lens All-Sky Survey (CLASS; \citealt{my03}). The black hole sample used for this study is described in Section 2. In Section 3, we discuss the flat-spectrum radio sources on the fundamental plane and use them in Section 4 to shed light on the beaming properties of the population. Finally in Section 5, we discuss our results and present the conclusions of this study.


\section{Sample Selection}

\subsection{Blazar sample}

\begin{figure}
\centering
\includegraphics[width=82.5mm]{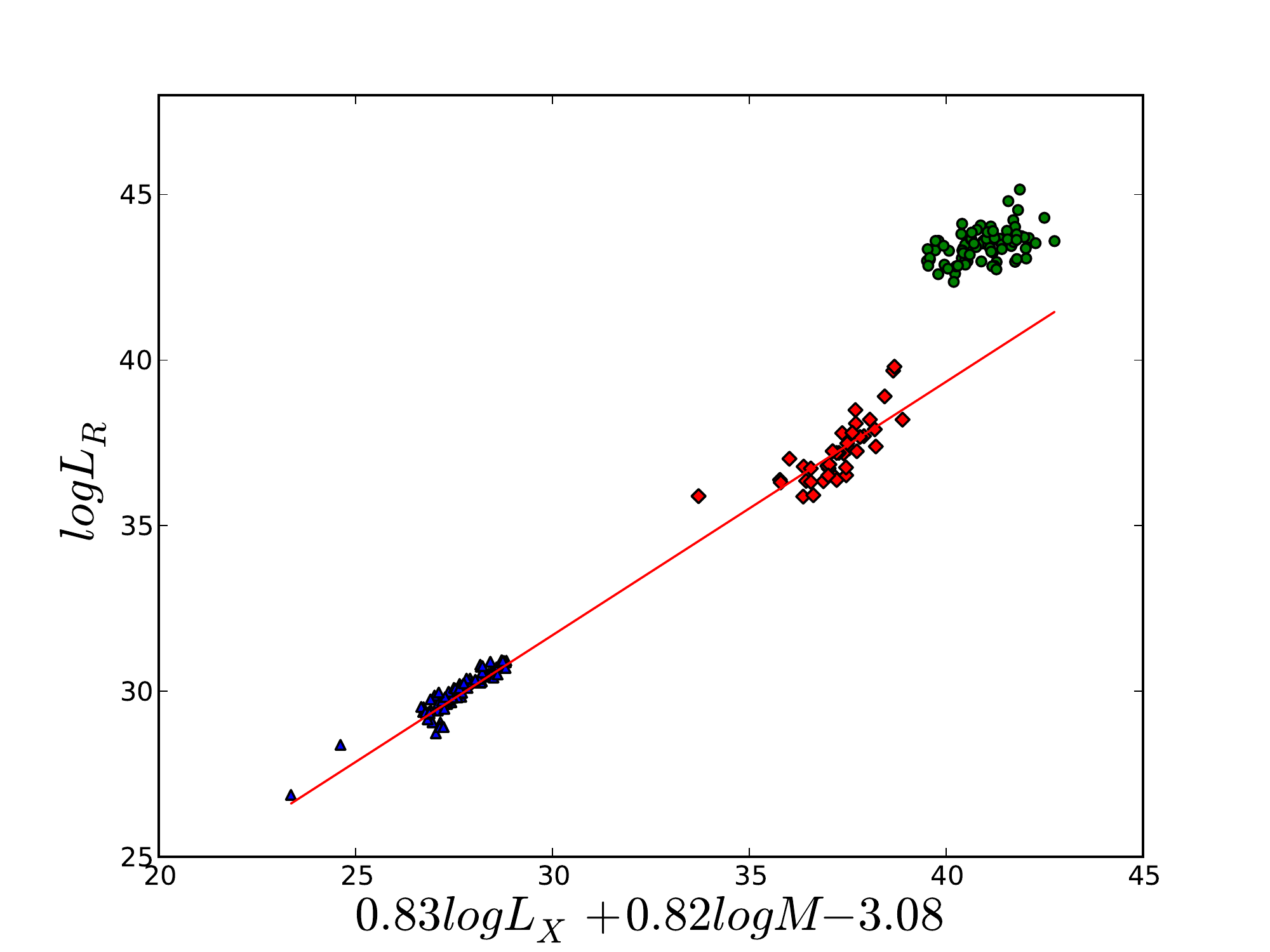}
\caption {Position of the blazars on the optical fundamental plane. Here, XRBs are in blue triangles, LLAGN in red diamonds and the VIPS Blazars are plotted as green circles. The red line is the projection of the best-fit plane using just the LLAGN sample. Luminosities are given in erg s$^{-1}$ while the masses are in the unit of solar mass.}
\label{fig:fp}
\end{figure}

The blazars used in this study are taken from the VLBA Imaging and Polarimetry Survey (VIPS; \citealt{helm07}) consisting of 1,127 sources with flat radio spectra. The parent sample for this survey is the Cosmic Lens All-Sky Survey (CLASS; \citealt{my03}) - a VLA survey of $\sim$12,100 flat-spectrum objects. CLASS is the largest and best studied statistical sample of radio-loud gravitationally lensed systems with compact flat radio spectrum. The majority of the CLASS sources are highly relativistically beamed. The parent population of the CLASS sources consist of radio-loud AGN, which are significantly brighter than the low-luminosity AGN (LLAGN) used to construct the optical FP, as demonstrated by their higher radio and [OIII] line luminosities (see Fig \ref{fig:fp}). They will then consist of FR I/II radio galaxies and radio-loud quasars, which have been shown to follow the X-ray fundamental plane of black hole activity if one takes beaming into account \citep[eg.][etc.]{m03,p12}.

VIPS has selected all CLASS sources on the survey area of the Sloan Digital Sky Survey (SDSS; \citealt{y00}), with an upper declination limit of $65^{\circ}$ and a lower declination limit of $15^{\circ}$, having flux densities greater than 85 mJy at 8.5 GHz, yielding a complete sample of 1,127 sources. Of these, 141 have already been observed at 5 GHz with the VLBA as part of the Caltech-Jordell Bank Flat spectrum survey (CJF; \citealt{t96}), 8 have been observed at 15 GHz as part of the Monitoring of Jets in AGN with VLBA Experiments project (MOJAVE; \citealt{lh05}), and 20 were observed for the VIPS pilot program at 5 and 15 GHz \citep{t05}. The remaining 958 sources were observed with the VLBA. 

For this study, we take only those sources for which we have available radio luminosity at 8.5 GHz (from CLASS survey taken with VLA at configuration A), black hole mass and [OIII] emission line luminosity at 5007\AA (from the catalog of quasar properties from SDSS DR7, \cite{s05}). The sources in this sample are dominated by FSRQs and BL Lacs. We would like to point out that the current method requires [OIII] line luminosities as tracer of accretion rate, and hence it does not work for extreme featureless BL Lacs. The optical FP works for both low and high excitation sources (LLAGN and seyferts). Hence the FSRQs are expected to follow the plane relation \citep[see also][]{m03}. In order to compare these sources with the optical FP, the 8.5 GHz fluxes were converted to 5 GHz fluxes using a flat radio spectra. In order to properly constrain the parameters of the proposed correlation, we exclude the upper limits from our final data set. But this sample may contain few soft state sources that can introduce additional uncertainties in the intrinsic properties. This gives us a final sample of 82 blazars. 

\subsection{Low Luminosity AGN in the Optical FP}

The LLAGN sample used in the optical fundamental plane of black hole activity \citep{s15} is extracted from the Palomar Spectroscopic Survey \citep{h95}, comprising the nuclear region of 486 bright nearby galaxies with the B-magnitude in Tycho photometric band $B_{T} < 12.5$ mag.

Radio luminosities, needed as estimate for the jet emission, have been taken from the high-resolution radio survey of all LLAGNs and AGNs in the Palomar sample, reported in \cite{n05} while the [OIII] line luminosities, used as a tracer of the accretion rate, have been taken from \cite{h97}, where spectroscopic properties and parameters for 418 AGN and their host galaxies are presented. Stellar velocity dispersions presented in \cite{ho09} using the same spectra, are used to derive the black hole masses using the M-$\sigma$ relation having an absolute scatter of 0.34 dex \citep{mf01,kh13}.

The fundamental plane parameters are obtained using directly observed [OIII] and radio fluxes, and inferred quantities like distances and masses. In order to give a physical meaning to the relation and to compare the LLAGN sample with the stellar mass XRBs, inferred black hole masses are used in the relationship instead of the directly observed stellar velocity dispersions. Moreover, a Kendall Tau Partial Correlation analysis was performed in \cite{s15} with distance as the third variable, and the FP relation was found to be real even after taking into account the large range of inferred distance. The chi-square confidence map for the parameters of the fundamental plane was found to be an elongated ellipse, showing that the parameters are coupled \citep{s15}. The uncertainties in the directly measured quantities like radio and [OIII] fluxes and the inferred quantities like distances and black hole masses are properly taken into account using the merit function while deriving the fundamental plane parameters. Restricting the final sample only to the galaxies which have available data in 15 GHz radio luminosity, [OIII] emission line luminosity and black hole mass; and after removing the upper limits; the sample size was reduced to 39 SMBHs.

\subsection{Stellar mass black holes in the Optical FP}

The stellar mass XRB sample used in the optical FP comprises the best-studied XRBs in the hard state : GX 339-4 (88 quasi-simultaneous radio and X-ray observation from \citealt{c13}), V404 Cyg (VLA radio and Chandra X-ray reported in in \citealt{c08}), XTE J1118+480 (from the compilation in \citealt{m03}) and A06200-00 (simultaneous Chandra X-ray and VLA radio observation reported in \citealt{g06}).

\section{Theoretical Sample}

\subsection{Relativistic beaming in blazars}

\begin{figure}
\centering
\includegraphics[width=82.5mm]{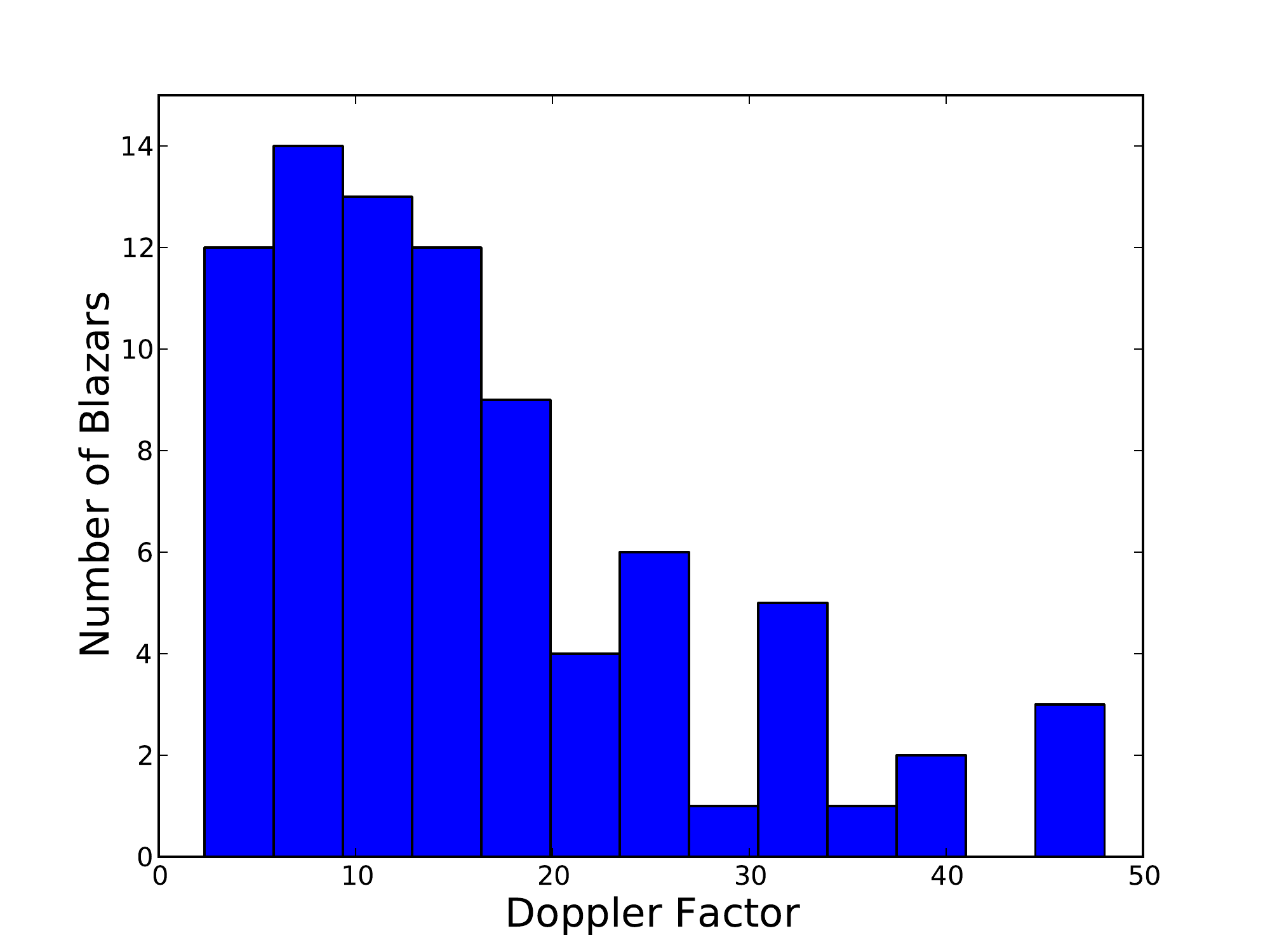}
\caption {Doppler distribution of the VIPS blazar sample}
\label{fig:df}
\end{figure}

All radio-bright active galactic nuclei have relativistic jets emitting synchrotron radiation. Due to relativistic effects, observed emission from these jets can be considerably brighter compared to the intrinsic luminosity. If L$_{int}$ is the intrinsic luminosity and L$_{obs}$ is the observed luminosity, then $L_{obs} = \delta^b L_{int}$, where $\delta$ is the kinematic Doppler factor given as
\begin{equation*}
\delta = \frac{1}{\Gamma(1-\beta Cos\theta)}
\end{equation*}

where $\beta$ is the bulk jet velocity in units of $c$, $\Gamma$ is the Lorentz factor defined as $(1-\beta^2)^{-1/2}$ and $\theta$ is the viewing angle of the jet with respect to our line-of-sight.

The exponent $b$ depends on various assumptions about jet structure, jet emission spectrum and the frequency at which the jet is observed. For a spectral index $\gamma$ defined as $S_{\nu}\propto\nu^{-\gamma}$, the boosting index $b$ can be approximated to be $b = 2+\gamma$ or $3+\gamma$ respectively in the case of a continuous jet or a jet with distinct blobs \citep{bk79, up95}. Blazars generally have flat spectra ($\gamma \sim 0$) and hence we take $b = 2$ for this study. The possible different choices of $b$ are discussed in Section 4.3.

\subsection{Blazars in previous fundamental planes}

Previously, relativistically beamed BL Lac objects were either excluded from the fundamental plane studies as the jet emission is heavily boosted \citep{m03}, or Doppler boosting was corrected assuming an average Doppler factor $\sim$ 7 \citep{f04}. In the latter case, it was seen that the X-ray/radio ratio is rather insensitive to the Doppler factor and the main effect was just a shift in the position of the sources along the correlation - the BL Lacs were pushed from the upper end of the correlation into the regime where FR I radio galaxies lie. This was theoretically expected as both X-ray and radio luminosities are subjected to relativistic beaming. Studying the blazars with the optical FP can give us more insight as unlike the X-ray and radio luminosities, the [OIII] emission line luminosity is assumed to be direction-independent and hence not affected by relativistic boosting.

\begin{figure}
\centering
\includegraphics[width=82.5mm]{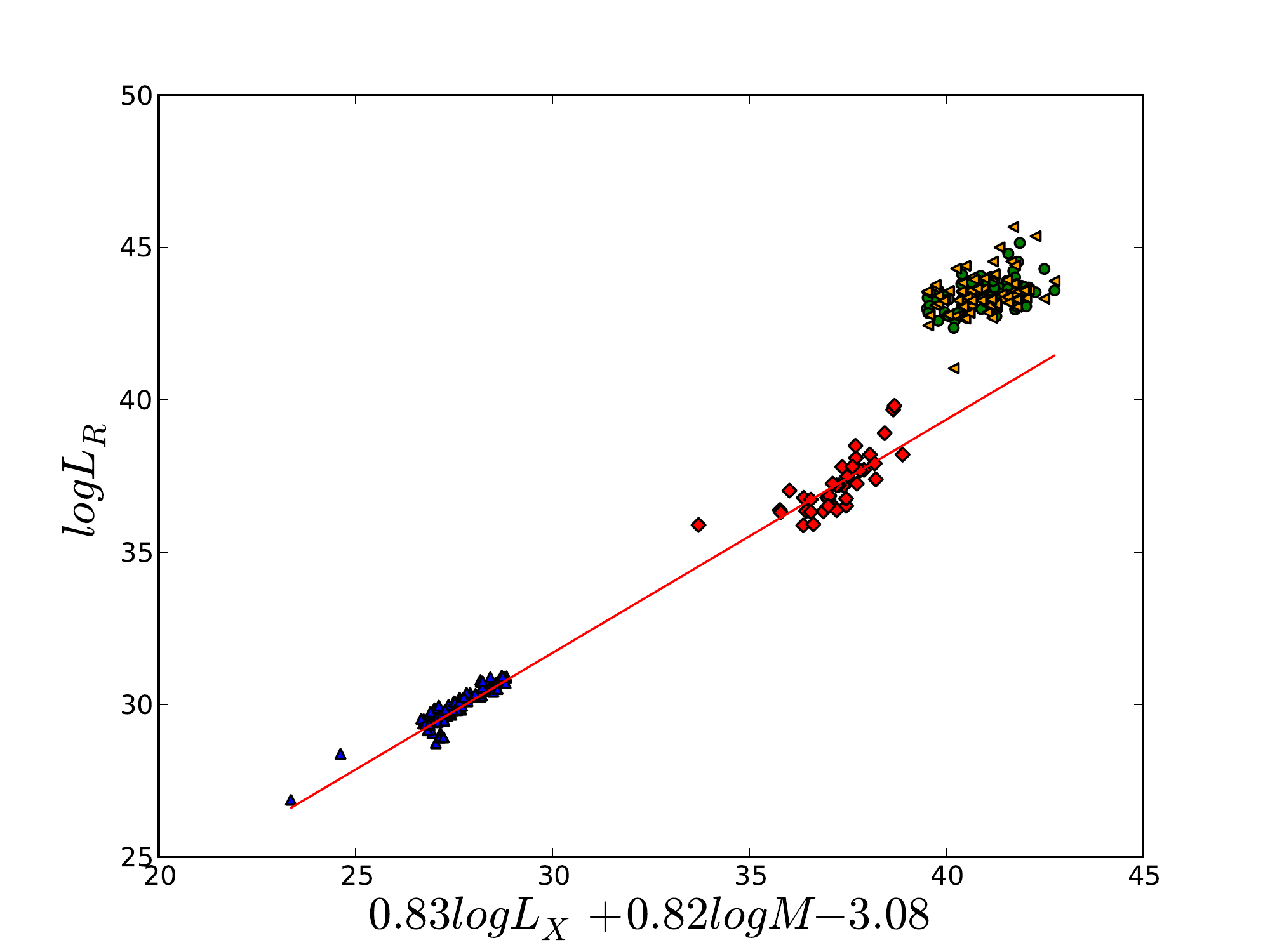}
\caption {VIPS observed blazars (green circles) and the simulated blazars (orange sided-triangles) on the optical fundamental plane. Luminosities are given in erg s$^{-1}$ while the masses are in the unit of solar mass.}
\label{fig:boo}
\end{figure}

\cite{s15} construct the optical FP by performing a multivariate correlation linear regression analysis on the SMBH sample. The relation was then extrapolated to lower black hole masses and the stellar mass black hole were consistent with the plane found with only the SMBHs. We now extrapolate the same relation to the higher black hole mass region and put the VIPS blazar data on the previous plane, without fitting them (see Fig \ref{fig:fp}). The blazars were found to have much higher radio luminosity compared to what was expected by the fundamental plane, which can be explained by relativistic beaming and the VIPS flux density limit of 85 mJy at 8.5 GHz.

This optical FP can be used to construct the Doppler factor distribution of the VIPS blazars (see Fig \ref{fig:df}). Given this Doppler factor distribution and the selection effects of the VIPS sample, we can constrain the Lorentz factor and viewing angle distributions for the sample.

\subsection{Simulating a blazar sample from optical FP}

We take the optical fundamental plane relation $\log L_{R} = 0.83 \log L_{OIII}  + 0.82 \log M$, and calculate the radio luminosities of the VIPS blazars using their [OIII] line luminosities and central black hole masses. These are the intrinsic radio luminosities of the sources, without the effect relativistic beaming.

\begin{figure}
\centering
\includegraphics[width=82.5mm]{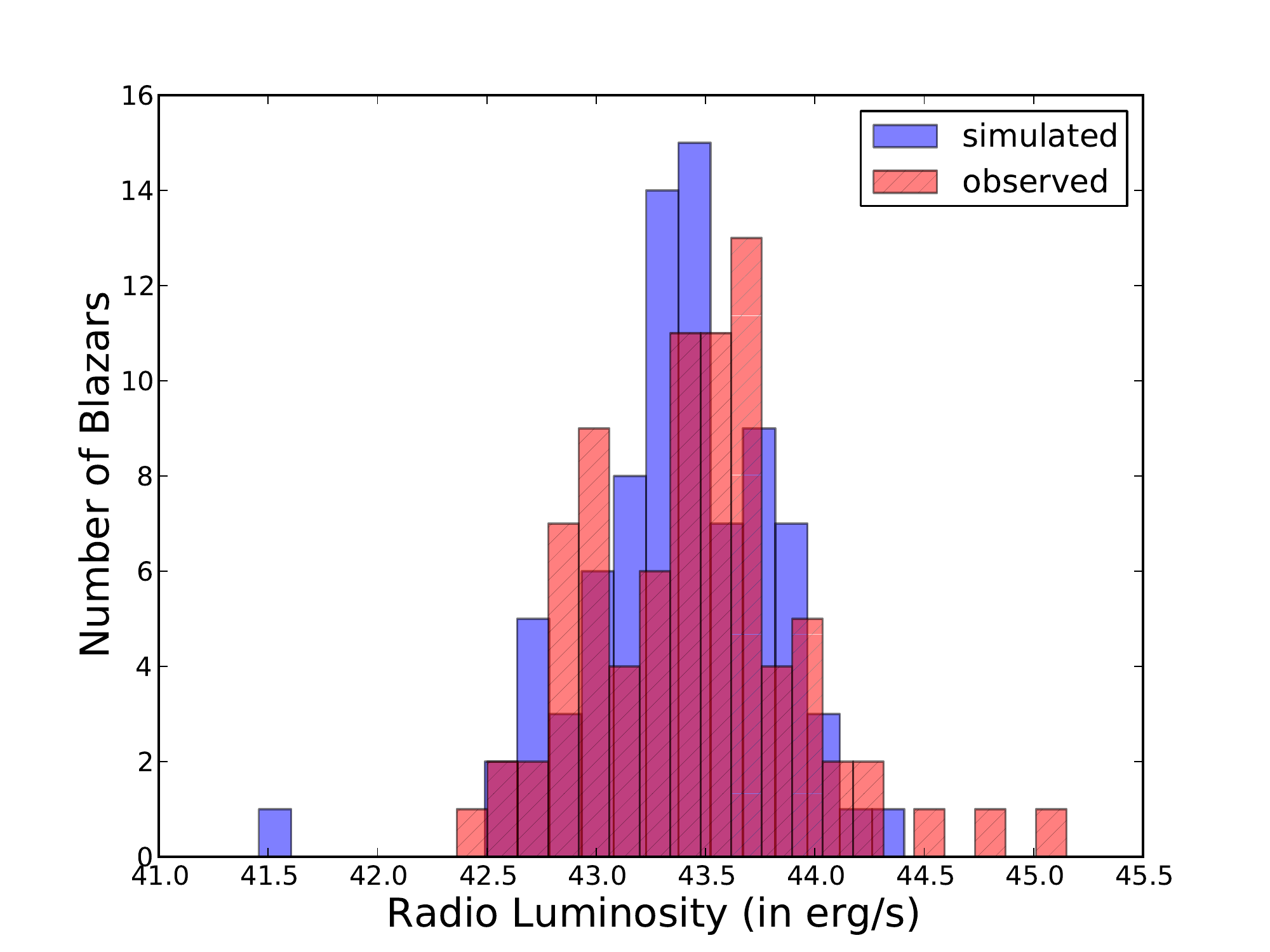}
\caption {Radio luminosity histograms : In blue are the FP-simulated blazars and in red (filled with dashed lines) are the blazars as observed by VIPS. The radio luminosity values are in erg s$^{-1}$, and are plotted in the logarithmic scale.}
\label{fig:hist}
\end{figure}

The intrinsic radio luminosities of the blazar sample are theoretically boosted using monte carlo simulation for a given viewing angle and Lorentz factor distribution. The resultant theoretically beamed radio luminosity distribution is then compared with the observed radio luminosity distribution obtained from the VIPS observations. The intrinsic scatter of the optical FP (0.35 dex) is added in the simulation to include the errors of the plane in the simulated sample.

The viewing angles were randomly chosen from an uniform distribution in cos $\theta$ for a $\theta$ range of $0^{\circ}$ to 30$^{\circ}$, assuming that the sources are randomly aligned with respect to the observer's line of sight. As explained in Section 4.2, the selection biases were later taken into account while simulating the blazars from the fundamental plane and the final $\theta$-distribution is found to be steeper at lower viewing angles. The Lorentz factor $\Gamma$s were taken from a bounded power-law parent distribution of the form $N(\Gamma) \propto \Gamma^{\alpha}$ as suggested by previous studies \citep[eg.][etc.]{pu92, lm97}. The use of other possible Lorentz factor distributions are discussed in Section 4.3. Moreover, in order to match the sensitivity of the simulated sample with the observed one, we incorporate the selection effect of the VIPS sample (radio flux-cut of 85 mJy at 8.5 GHz) in the monte carlo simulation.

Once the theoretically simulated sample is constructed, we compare it with the VIPS observed blazar sample. As seen in Fig \ref{fig:boo}, the simulated sample occupies the same space as the VIPS observed sample.

\subsection{Statistical test on the simulated sample}

We compare the histograms of the radio luminosity distribution of the VIPS observed blazars with the distribution of the theoretical sample simulated from the fundamental plane in Fig \ref{fig:hist}. A Kolmogorov-Smirnov (KS) test was performed to quantify these similarities in the radio luminosity distributions of the two samples. If the KS statistic is small or the p-value is high, then we cannot reject the null hypothesis that the underlying distributions of the two samples are the same.

In order to statistically test if the simulated sample matches with the radio luminosity distribution of the VIPS sample, we assume a power law Lorentz factor distribution $N(\Gamma) \propto \Gamma^{-2}$, bounded between 1 and 30, as our initial guess. A KS test done on the VIPS observed sample and a simulated sample, gives a KS statistic value of 0.12$\pm$0.03 and a p-value of 0.51$\pm$0.23, thereby showing that the underlying radio luminosity distributions of both the samples are in agreement with being similar.

\section{Results}

\subsection{Lorentz factor distribution}

\begin{figure}
\centering
    \includegraphics[width=81mm]{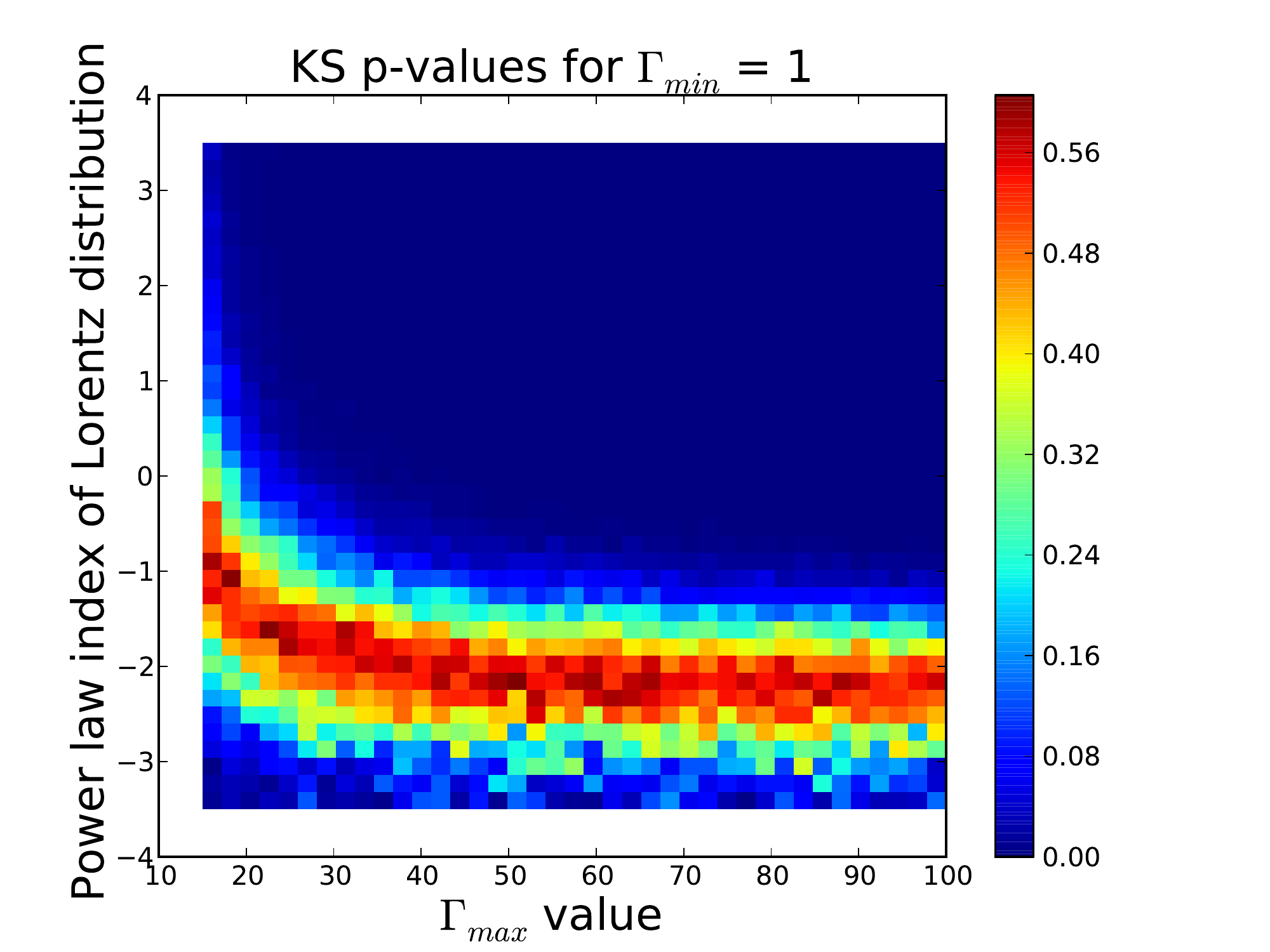}
 \caption{For $\Gamma_{min}$ of 1, color bar diagrams depicting the KS p-values for different combinations of $\Gamma_{max}$ and $\alpha$. For higher KS statistic or smaller p-value, we can reject the hypothesis that the distributions of the two samples are the same.}
\label{fig:color}
\end{figure}

The KS statistic values and the p-values obtained from the simulated sample can be used to constrain the Lorentz factor distribution of the VIPS blazar sample.

As the first step, we fix the minimum Lorentz factor ($\Gamma_{min}$) of the VIPS sample, and compare the KS p-values for different combinations of maximum Lorentz factor ($\Gamma_{max}$) and the power law index of the $\Gamma$-distribution (i.e. $\alpha$ in $N(\Gamma) \propto \Gamma^{\alpha}$). For higher KS statistics or smaller p-values, we can reject the hypothesis that the distributions of the theoretically simulated sample and the observed sample are the same. Color-bar diagrams depicting the KS values for different combinations of $\Gamma_{max}$ and $\alpha$ can be seen in Fig \ref{fig:color}. For the blue parts of the plot i.e. when the p-value is small ($<$ 0.15), we can reject our null hypothesis. Hence, we can constrain a very thin range of possible Lorentz factor distributions, for which the theoretically simulated sample can match the VIPS observed ones. This gives a rough idea about the possible values of $\Gamma_{max}$ and the power-law exponent $\alpha$ for a given $\Gamma_{min}$.

As evident from Fig \ref{fig:color}, for a fixed $\Gamma_{min} = 1$, the shape of the $\Gamma$ distribution shows a steep evolution for small $\Gamma_{max}$ values, but it gets almost constant for any choice of $\Gamma_{max}$ higher than 40. Previous observational studies of relativistic jet Lorentz factors have reported $\Gamma_{max}$ values of 40 or more \citep[eg.][etc.]{m06,h09,s10,a12}. For example, the MOJAVE survey of compact, highly beamed radio-loud AGNs analyzed 127 jets and used the apparent velocities of the jet components to report a Lorentz factor distribution in the parent population that range up to $\sim$50 \citep{l09}.

The maximum possible Lorentz factor in our distribution needs to be more than what is observed. It is clear from Fig \ref{fig:color} that for any choice of $\Gamma_{max}$ higher than 40, there is no considerable evolution in the value of $\alpha$ . Hence, to evaluate the Lorentz factor distribution of the VIPS blazar sample in more detail, we can safely fix the $\Gamma_{max}$ value at 40, and the resultant shape of the distribution can be considered to be true for other values of $\Gamma_{max} > 40$.

Then, for a given $\Gamma_{min}$ and $\Gamma_{max}$, it is possible to constrain the shape of the Lorentz factor distribution statistically by performing a KS test between the theoretically simulated sample from the optical FP and the observed VIPS sample.

\begin{figure}
\centering
\includegraphics[width=80mm]{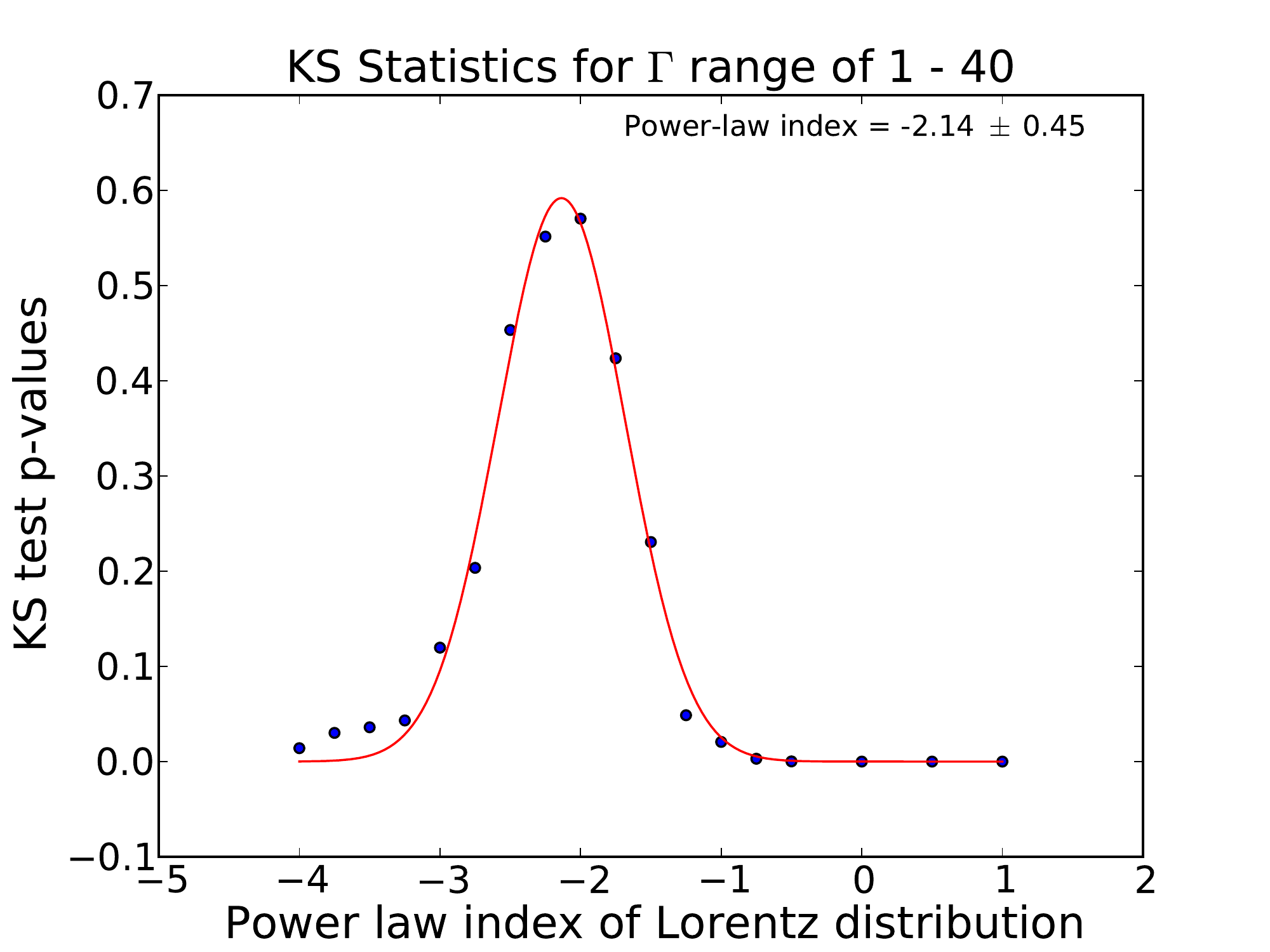}
\caption {KS statistics for different power-law exponents depicting the blazar $\Gamma$ distributions for a fixed $\Gamma$ range of 1-40 and a $\theta$ range of 0-30, showing a best-fit value for  $-2.1 \pm 0.4$.}
\label{fig:gamdist}
\end{figure}

With a fixed $\Gamma$ range of 1-40 and $\theta$ range of 0-30, we use Monte Carlo simulation to create different theoretically boosted samples for various power-law exponents ($\alpha$ of $N(\Gamma) \propto \Gamma^{\alpha}$). We compare all these theoretical samples with the VIPS observed sample, compute their KS values and find that the highest KS p-values are obtained for a power law index of -2.1. In Fig \ref{fig:gamdist}, we show the KS values obtained for different power-law exponents. From the KS p-values obtained, we can reject the null hypothesis of the two samples being similar for all the $\alpha$ values outside the range of $\sim$-1 to  $\sim$-3. We can statistically constrain the Lorentz factor distribution for a $\Gamma$ range of 1-40 to be a power law $N(\Gamma) \propto \Gamma^{-2.1 \pm 0.4}$ .

\subsection{Viewing angle distribution}

The parent viewing angle ($\theta$) distribution was taken as an uniform distribution ranging from $0^{\circ}$ to 30$^{\circ}$, properly weighed by the 3D spherical distribution function.

But as mentioned in Section 3.2, while boosting the simulated blazars from the fundamental plane, we put in the selection effect of the VIPS sample (radio flux-cut of 85 mJy at 8.5 GHz) to match the sensitivity of the simulated sample with the observed one. This ensures that only those $\theta$s are selected, for which the theoretical blazar sample from the optical FP can be boosted enough to clear the radio flux cut-off needed for the VIPS sample.

\begin{figure}
\centering
\includegraphics[width=80mm]{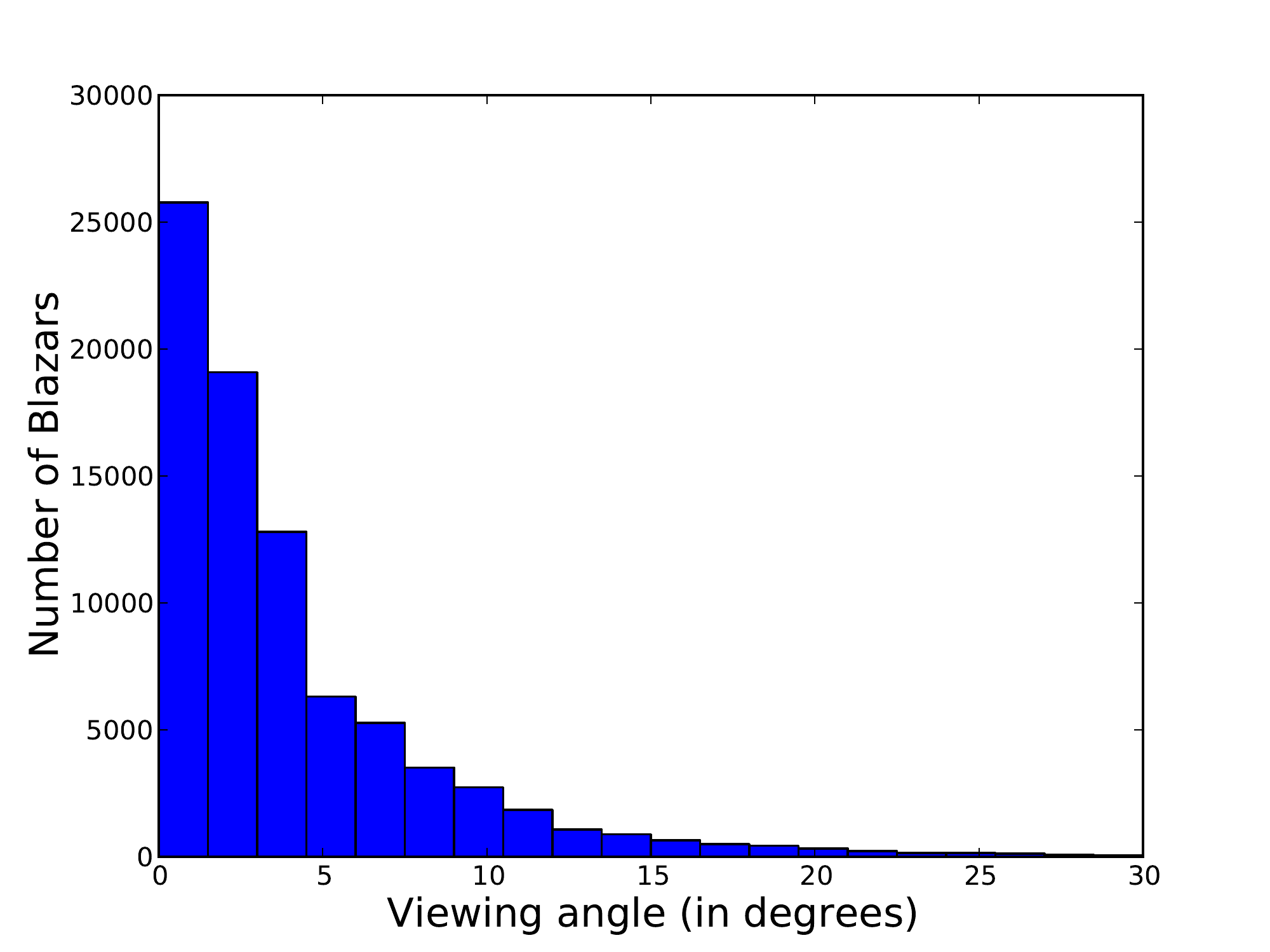}
\caption {Distribution of viewing angles for the blazars, after bootstrapping the theoretical sample simulated from the fundamental plane.}
\label{fig:theta}
\end{figure}

In Fig \ref{fig:theta}, we show the theta distribution obtained from simulations of various bootstrapped VIPS blazar sample. As evident from Fig \ref{fig:theta}, this condition on sensitivity results in the final $\theta$-distribution to be steepest at lower viewing angles, which flattens out as the viewing angle increases. Hence, we conclude that most of the sources in a blazar population have a viewing angle smaller than $\sim5^{\circ}$, which agrees with previous studies \citep[eg.][etc.]{a12}.

\begin{figure*}
  \includegraphics[height=43.5mm]{Paper_presentation_140_030_1_index2.pdf}
      \includegraphics[height=43.5mm]{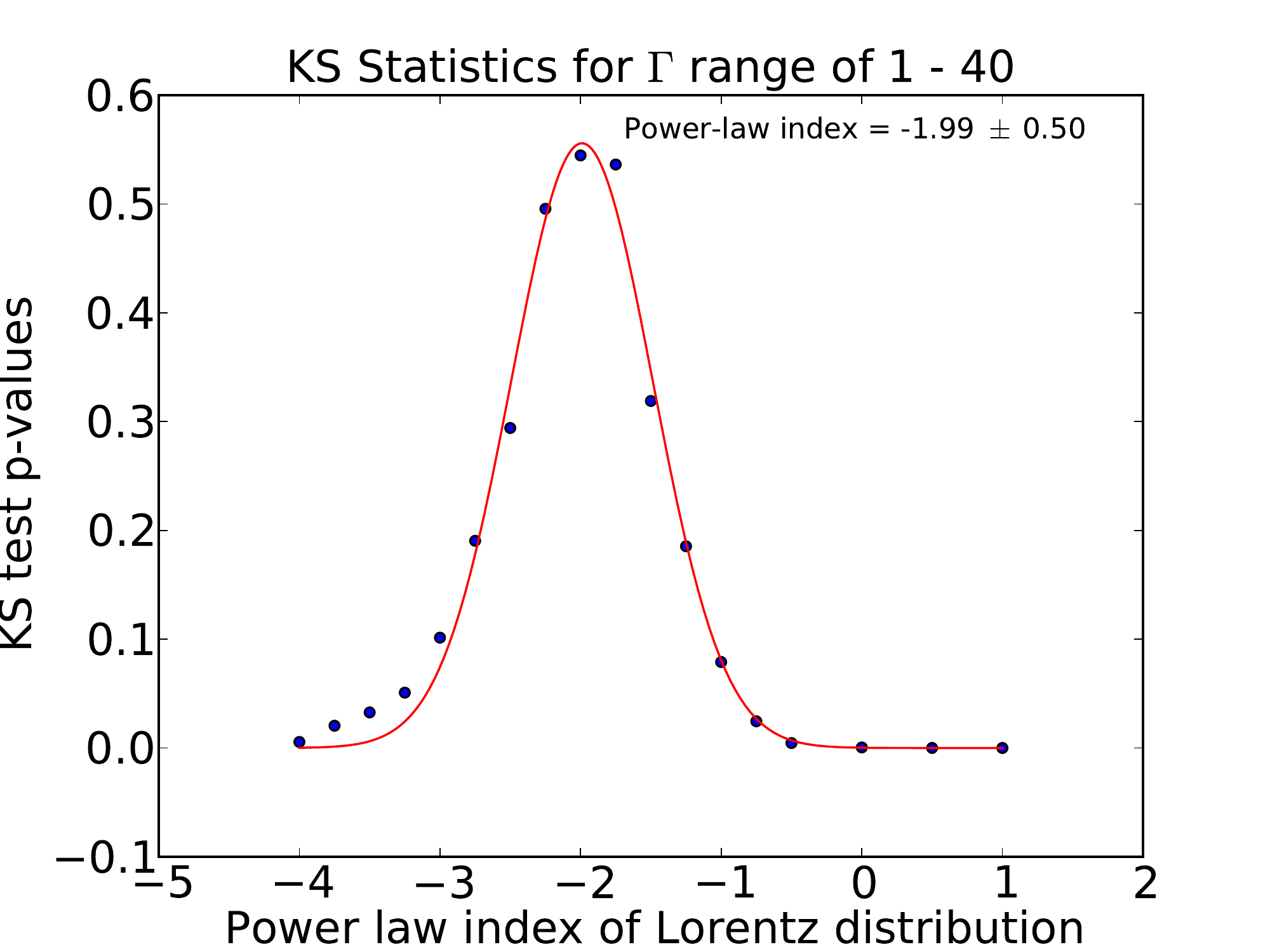}
    \includegraphics[height=43.5mm]{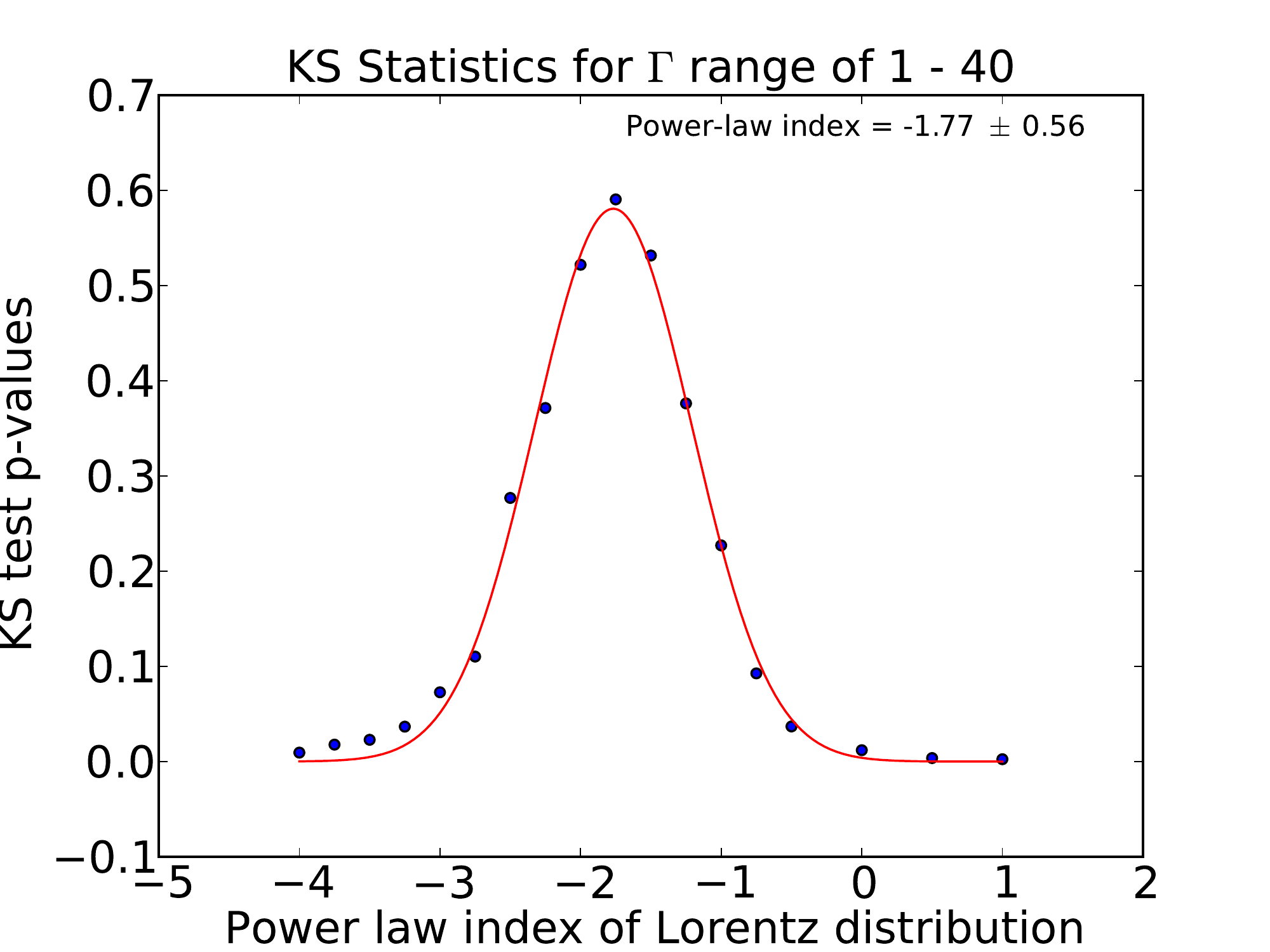}
 \caption{KS p-values for different power-law indices of $\Gamma$ distributions, for a fixed $\Gamma$ range of 1-40 with different $\theta_{min}$ values of $0^{\circ}$, $1^{\circ}$ and $2^{\circ}$, respectively.}
\label{fig:thegam}
\end{figure*}

\subsection{Dependence of the result on input parameters}

It is important to examine if the derived distribution of Lorentz factors depends on the input parameters used in the analysis. Here we discuss how different choices of input parameters affect the agreement between the VIPS observed sample and the theoretical sample simulated from the optical fundamental plane.

\subsubsection{$\Gamma$ range}

The range of possible Lorentz factors of flat-spectrum sources is still uncertain. For our analysis, we have used a $\Gamma$ range of 1-40, though this range can easily be much larger. But as shown in the KS p-value plot of Fig 5, for a given $\Gamma_{min}$, different choices of $\Gamma_{max}$ do not change the final result considerably. A detailed analysis done for a $\theta$ range of $0^{\circ} - 30^{\circ}$ with a $\Gamma_{min}$ = 1 shows that the power-law exponent of $\Gamma$ distributions can be constrained as $-2.14 \pm 0.45$, $-2.14 \pm 0.44$ and $-2.18 \pm 0.42$ for $\Gamma_{max}$ values of 40, 60 and 90, respectively. So the details of the distribution does not change much with different $\Gamma_{max}$ values. Also, having a $\Gamma_{min}$ value larger than 3 can be ruled out for our sample, as for no choice of $\Gamma_{max}$, a KS p-value higher than 0.3 could be obtained for a $\Gamma_{min}$ value higher than 3. We need to include blazar sources having Lorentz factor values lower than 3 to properly constrain the Lorentz factor distribution of the population using our method. Hence, although blazars are considered to be highly relativistic sources, our analysis indicates that the possibility of having sources with $\Gamma$ values less than 3, can not be ruled out for the given sample. This suggests that the sample we have used may not be completely pure blazars and may contain few sources which are flat spectrum radio sources that are not highly relativistically beamed.

\subsubsection{$\theta$ range}

As seen in Fig \ref{fig:theta}, the only assumptions on the parent viewing angle distribution, that can change the boosting parameters considerably, is the choice of the $\theta_{min}$ value. In this study, we have taken $\theta_{min}$ to be $0^{\circ}$. In Fig \ref{fig:thegam}, we constrain the power-law exponents of $\Gamma$ distributions for a fixed $\Gamma$ range of 1-40, with $\theta_{min}$ values of $0^{\circ}$, $1^{\circ}$ and $2^{\circ}$ to be $-2.1 \pm 0.4$, $-1.99 \pm 0.5$ and $-1.77 \pm 0.56$, respectively.

From the constrained power-law indices, we see that using a higher $\theta_{min}$ results in a lower value of the power-law index with a higher standard deviation. That is expected as standard models of relativistic jets predict that the opening angle of the jet should be inversely proportional to the Lorentz factor \citep[eg.][]{bk79}. Hence introducing a higher $\theta_{min}$ cut will effectively lower the $\Gamma_{max}$ value and suppress higher $\Gamma$ factors in the Lorentz factor distribution. For example, $\theta_{min}$ of $2^{\circ}$ implies a $\Gamma_{max} \sim 30$, with a corresponding suppression of $\Gamma$ factors near the upper limit of the distribution, i.e. around $\Gamma \sim$30.

\subsubsection{Boosting index value $b$}

VIPS, the parent sample of this study, is selected to consist of flat-spectrum objects with a spectral index $\gamma>$ -0.5 (between 4.85 GHz and a lower frequency). Theoretically, for a relativistically beamed source, the boosting index is $b = 2+\gamma$ for a continuous jet and $3+\gamma$ for a jet with distinct blobs \citep{lb85}. As the fundamental plane is theoretically based on a continuous jet, for our analysis we use the case of $b = 2+\gamma$, where $\gamma$ is taken to be 0 for the blazars.

We see that increasing the boosting index $b$, results in a higher power-law index of the Lorentz factor distribution, with a bigger uncertainty. The best-fit $\Gamma$-distribution using a continuous jet assumption with flat spectra ($\gamma =$ 0) was found to be $N(\Gamma) \propto \Gamma^{-2.1 \pm 0.4}$ for the $\Gamma$ range of 1 to 40. This power-law index increases gradually from $-2.1 \pm 0.4$ to $-4.0 \pm 1.0$, as the boosting index is increased from 2.0 (continuous jet) to 3.0 (ballistic jet).

\subsubsection{Different $\Gamma$ distribution possibilities}

Previous studies have suggested and explored a power-law distribution for Lorentz factors \citep[eg.][etc.]{pu92,j05}. In this analysis, we have tried to constrain the power-law index for this distribution. We have also checked for other simple and possible forms of $\Gamma$ distributions like uniform and delta function distributions in the VIPS sample.

A two sample KS test analysis between the observed VIPS sample and a theoretical sample simulated from the fundamental plane using an uniform $\Gamma$ distribution of 1-40 gave a KS p-value of 0.25$\pm$0.0001, with the KS p-value decreasing even more with an increase in the $\Gamma$-range to values such as a KS p-value of $6 \times10^{-5}\pm 10^{-4}$ for an uniform $\Gamma$ distribution of 1-60. Hence we can safely rule out the possibility of having a uniform Gamma-distribution for the VIPS blazars.

Similarly, we also analyzed the agreement of the simulated and the observed sample for delta function distributions, i.e. single values of Lorentz factors. It was seen that the Lorentz factor delta distribution is required to center around a $\Gamma$ value more than 23, for all the simulated blazars to clear the VIPS sample flux cut-off of 85 mJy at 8.5 GHz. And even these accepted Lorentz factors result in very low KS p-values (KS p-value of 0.0003 for a delta distribution of $\Gamma$=25, with the p-value decreasing even more with an increase in the $\Gamma$ value, e.g. a KS p-value of $3 \times10^{-9}$ for a $\Gamma$=30). Therefore, for the theoretically beamed sample to agree with the observed one, we can rule out a delta function distribution for the VIPS blazar Lorentz factors.\\

Hence, we can summarize that although the result can slightly change with some of the assumptions and input parameters like the boosting index value, the agreement between the theoretical and the observed sample is quite robust for a power law form of $\Gamma$-distribution with indices $-2.1 \pm 0.4$.
 
\section{Discussion and conclusion}

In this paper, we present a new and independent method of calculating the Lorentz factor distribution of a blazar population using the optical fundamental plane of black hole activity. We use the VIPS blazar sample and find the bulk Lorentz factor distribution to be in the form of a power law given as $N(\Gamma) \propto \Gamma^{-2.1 \pm 0.4}$, with a $\Gamma$ range of 1 to 40.

The optical fundamental plane is an efficient and insightful platform to study high-luminosity blazars, since, unlike the X-ray and radio luminosities, the [OIII] emission line luminosity of a blazar is direction-independent and hence not affected by relativistic boosting. Another advantage of using the optical FP is that the blazars are high mass sources and hence the synchrotron cut-off of the jet spectrum can already be below the X-ray band, specially for the low-frequency-peaked BL Lacs. Thus, the observed X-ray luminosity will be dominated by synchrotron self-Compton emission and hence can not be used as a proxy for the accretion rate.

We take VLBI radio luminosities for the blazars observed in the VIPS survey and use the theoretical predictions from the optical FP to find a blazar Lorentz factor distribution in the form of a power law. A power law form of the Lorentz factor distribution has been explored earlier. \cite{pu92} had found a Lorentz factor distribution in the form of $N(\Gamma) \propto \Gamma^{-2.3}$, while \cite{lm97} had used a distribution of the form $N(\Gamma) \propto \Gamma^{\alpha}$, where $-1.5 \lesssim \alpha \lesssim -1.75$.

Here, we constrain the blazar Lorentz factor distribution along with its uncertainties, within the error range of the previous studies, to be a power law of the form $N(\Gamma) \propto \Gamma^{-2.1 \pm 0.4}$ in the $\Gamma$ range of 1 to 40. This Lorentz factor distribution of blazar population can be used in relativistic beaming models of blazar sources to properly constrain the physics of jet launching region and estimate their underlying intrinsic physical properties. The $\Gamma$-distribution presented in this paper has been obtained for the VIPS sample of blazars, which lie in the declination range of $15^{\circ}$-$65^{\circ}$ and have a flux density of 85 mJy or higher at 8.4 GHz. The finding should hold for the whole parent population of the CLASS sample, which consists mainly of radio galaxies and flat spectrum radio quasars. With a bigger, statistically complete sample of blazars and further refinement of the optical fundamental plane, it will be possible to place tighter constrains on the shape of this distribution.

\section*{ACKNOWLEDGEMENTS}
The authors acknowledge funding from an NWO VIDI grant under number 639.042.218.


\bsp

\label{lastpage}

\end{document}